\documentclass{llncs}

\usepackage{latexsym}
\usepackage{amsmath}
\usepackage{amssymb}
\usepackage{appendix}

\newcommand{\mon}{\textsc{MOn-1qfa}}
\newcommand{\mons}{\textsc{MOn-1qfa}s}
\newcommand{\lmo}{\textbf{LMO}}

\newcommand{\lipt}{\text{liId}\textbf{PT}}
\newcommand{\id}{\text{liId}}

\newcommand{\C}{\mathbb{C}}
\newcommand{\0}{\bf{0}}

\newcommand{\R}{\mathbf{R}}
\newcommand{\J}{\mathbf{J}}
\newcommand{\bg}{\mathbf{BG}}

\begin{document}

\title{Algebraic Characterization of the Class of Languages recognized by Measure Only Quantum Automata \\ \hspace{1pt} \\ (Extended Abstract)
}
\author{Carlo Comin\inst{1,2} \and Maria Paola Bianchi\inst{1}}
\institute{ \textsuperscript{1}Dipartimento di Informatica, Universit\`a degli Studi di Milano, Milano, Italy
\textsuperscript{2}ESTECO, AREA di Ricerca Science Park, Padriciano 99, Trieste, Italy
 \email{carlo.comin.86@gmail.com, bianchi@di.unimi.it} }
\maketitle
\begin{abstract}
We study a model of one-way quantum automaton where only measurement operations are allowed (\mon). 
We give an algebraic characterization of $\lmo(\Sigma)$, showing that
 the syntactic monoids of the languages in $\lmo(\Sigma)$ are exactly the literal pseudovariety of $J$-trivial literally idempotent monoids, where $J$ is the Green's relation determined by two-sided ideals.
We also prove that $\lmo(\Sigma)$ coincides with the literal variety of literally idempotent piecewise testable regular languages. 
This allows us to prove the existence of a polynomial time algorithm for deciding whether a regular language belongs to $\lmo(\Sigma)$.
\end{abstract}

%
%
%
\subsubsection*{Introduction and preliminaries.}
This paper gives a characterization of the class of languages recognized by a model of quantum automata, by using tools from algebraic theory, in particular, varieties of languages.
Many models of one-way quantum finite automata are present in the literature: the oldest is the Measure-Once model \cite{BC01,BP02}, characterized by unitary evolution operators and a single measurement performed at the end of the computation. On the contrary, in other models, evolutions and measurements alternate along the computation \cite{Aal06,KW97}. The model we study is the Measure-Only Quantum Automaton (\mon), introduced in \cite{BMP10}, in which we allow only measurement operations, not evolution.
All these quantum models are generalized by Quantum Automata with Control Language \cite{BMP03}.

A \mon\ over the alphabet $\Sigma$ is a tuple of the form 
$
A = \langle \Sigma \cup \{\#\},$ $(O_c)_{c\in
  \Sigma\cup\{\#\}}, \pi_0, F \rangle
$. 
The complex $m$-dimensional vector $\pi_0\in\C^{1\times m}$, with unitary norm $||\pi_0||=1$, 
is called the quantum initial state of $A$.
For every $c\in\Sigma$, $O_c\in\C^{m\times m}$
is (the representative matrix of) an idempotent Hermitian operator and denotes an observable.
The subset $F\subseteq V(O_{\#})$ of the eigenvalues of $O_{\#}$ is called the spectrum of the quantum final accepting states of $A$.\mbox{}\\
The computation dynamics of automaton $A$ is carried out in the following way: 
let $x=x_1\ldots x_n\in\Sigma^*$, suppose we start from $\pi_0$, then $A$ measures the system with cascade observables 
$O_{x_1}, \ldots, O_{x_n}$ (by applying the associated orthogonal projectors) 
and then performs the final measure with the end-word observable $O_{\#}$, that is the observable of the final accepting states $F$ of $A$.\\
This last measure returns, as a result, an eigenvalue
$r\in V(O_{\#})$, if $r\in F$ then we say that the automaton $A$ accepts the word $x\in\Sigma^*$, otherwise that $A$ does not accept it. 
What is remarkable in this computation dynamics is the probability $p_A(x)$ that $A$ accepts $x=x_1\cdots x_n$.
In the specific case of \mons\, it turns out to be of some interest to express $p_A(x)$ using the well-known formalism of quantum density matrices.
We say that a language $L$ is recognized by $A$ with isolated cut point $\lambda$ iff for all $x\in\Sigma^*$\, $p_A(x)>\lambda \Leftrightarrow x\in L$ and there exists a constant value $\delta>0$ such that $|p_A(x)-\lambda|\geq\delta$.
%
%
%

We now recall some general definitions and results from the algebraic theory of automata and formal languages.
For more details, we refer the reader to, e.g. \cite{Ei76,Pi86}.
Let $L$ be a regular language and let $\langle \Sigma, Q, \delta, q_0,
F \rangle$ be the minimal deterministic automaton recognizing $L$. For a word $w=\sigma_1\cdots \sigma_n\in\Sigma^*$, we define its variation as 
$\text{Var}_L(w) =  \#\{0\leq k < n \mid \delta(\sigma_1\cdots \sigma_k)\neq \delta(\sigma_1\cdots \sigma_{k+1})\}$.
We say that $L$ has finite variation iff $ \text{sup}_{x\in\Sigma^*} \text{Var}_L(x) < \infty $.
Using results and similar techniques as in \cite{BMP10}, it is not difficult to show  
that the class $\lmo(\Sigma)$ of languages recognized by a \mon\ with isolated cut point is a boolean algebra of regular languages in $\Sigma^*$ with finite variation.
We say that a language $L\in\Sigma^*$ is literally idempotent iff for all $x,y\in\Sigma^*$ and $a\in\Sigma$, $xa^2y\in L\Leftrightarrow xay\in L$; we say that $L$ is literally idempotent piecewise testable if and only if it lies
in the boolean closure
of the following class of languages: 
$\Sigma^* a_1 \Sigma^* a_2 \Sigma^*\cdots \Sigma^* a_k \Sigma^*$, for $a_1, a_2, \ldots, a_k\in\Sigma$ and $a_1\neq \cdots \neq a_k$. 
We denote by $\id$ the class of literally idempotent languages and by $\lipt$ the class of literally idempotent piecewise testable languages.
For any language $L$, we call $M(L)$ its syntactic monoid.
We say that a class of finite monoids $\mathbf{A}$ is a (literal) pseudovariety if and only if it is closed under (literal) substructures, homomorphic images and finite direct products, \cite{KP08}.
Let $\mathbf{A}$ be a class of monoids and let $\Sigma$ be an alphabet. We denote by $V_{\Sigma}(\mathbf{A})$ 
the class of regular languages on $\Sigma$ having syntactic monoid in $\mathbf{A}$. 
Let $L, R$ and $J$ be the Green's relations determined by left, right and two-sided ideals, respectively.
In this paper we denote by $\R$ the pseudovariety of $R$-trivial finite monoids and by $\J$ the pseudovariety of $J$-trivial finite monoids.
We also define $\overline\J$ as the literal pseudovariety of $J$-trivial syntactic monoids $M(L)$ such that the associated morphism $\phi_L:\Sigma^*\rightarrow M(L)$ satisfies the literal idempotent condition $\phi_L(\sigma)\phi_L(\sigma)=\phi_L(\sigma)$, for every $\sigma\in\Sigma$.
We say that a class of regular languages $V:\Sigma\rightarrow 2^{\Sigma^*}$ is a $*$-variety of \emph{Eilenberg} if $V(\Sigma)$ is closed under boolean operations, right and left quotient, and inverse homomorphism. Replacing closure under inverse homomorphism by closure under inverse literal homomorphism, we get the notion of literal variety of languages.
%
%
%
A fundamental result is due to Eilenberg, who showed that there exists a bijection $V_\Sigma$ from the psuedovarieties of monoids and the $*$-varieties of Eilenberg of formal languages \cite{Pi86}.
In \cite{KP08}, Kl\'ima and Pol\'ak showed the following
\begin{theorem}\label{thm:klimapolak} Let $L\subseteq\Sigma^*$. 
It holds that  
$L\in\lipt$ if and only if 
$L\in V_{\Sigma}(\mathbf{J})\cap \id(\Sigma)$ if and only if  $L\in V_{\Sigma}\left(\overline{\J}\right)$.
\end{theorem}
%
%
%
%
%
%
\subsubsection*{Results.}
We give a direct proof that the class of finite variation regular languages is a $*$-variety of Eilenberg.
Moreover, we observe that a regular language $L$ has finite variation if and only if its syntactic monoid is $R$-trivial.
We proceed further on with our analysis by showing that the class of \mon\ over $\Sigma$ is in fact a sub-class of Latvian Automata. This class of 
automata has been fully characterized algebraically in \cite{Aal06} as the class of automata recognizing exactly regular languages having syntactic monoids in the class $\bg$ of block groups. 
\begin{theorem}
Let $A$ be a \mon\ on $\Sigma$ and let $L_A$ be a language recognized by $A$ with cut-point $\lambda$ isolated by $\delta$. 
Then there exists a Latvian automaton $A'$ recognizing 
$L_{A'}=L_{A}$ with cut-point $\lambda' = \frac{1}{2}$ isolated by $\delta' = \frac{\delta}{2\cdot \text{max}(\lambda, 1-\lambda)}$.
\end{theorem}
This directly implies that $\lmo(\Sigma)\subseteq V_{\Sigma}(\bg)$.
%
%
%

Combining our analysis with the results of \cite{Aal06} on block groups syntactic monoids and the results of \cite{BMP10} 
on finite variation languages, we prove the following
\begin{theorem}
Let $L\in\lmo(\Sigma)$ be a language recognized by some \mon\ with isolated cutpoint. Then its syntactic monoid $M(L)$ 
is an $R$-trivial block group, formally speaking $ M(L) \in \bg\cap \R $.
\end{theorem}
Since an $R$-trivial block group is also $J$-trivial, and since $\lmo(\Sigma)$ is a boolean algebra, we have $\lmo(\Sigma)\subseteq V_\Sigma(\J)$. This, together with Theorem \ref{thm:klimapolak} and the fact that languages in $\lmo(\Sigma)$ are literally idempotent, leads to the following
\begin{theorem}
$\lmo(\Sigma)\subseteq V_\Sigma(\overline\J)$.
\end{theorem}
%
%
%

We now show how languages in \lipt\ can be recognized by \mons.
Consider the language 
$L[a_1, \ldots, a_k] = \Sigma^*a_1\Sigma^*\cdots\Sigma^*a_k\Sigma^*$,
where $a_1,\ldots, a_k \in\Sigma$, $a_i \neq a_{i+1}$ for $1\leq i< k$, and let $S = \{a_1, \ldots, a_k\}$.
For every $\alpha \in S$, let $\#\alpha$ be the number of times that $\alpha$ appears as a letter in the word $a_1a_2\cdots a_k$.
Let $j^{(\alpha)}_1 < j^{(\alpha)}_2 < \cdots < j^{(\alpha)}_{\#\alpha}$ be all the indexes such that $\alpha =
a_{j^{(\alpha)}_1} = \ldots = a_{j^{(\alpha)}_{\#\alpha}}$. We define, for every $\alpha\in S$, two orthogonal projectors of dimension $(k+1)\times (k+1)$: the up operator $P^{(k)}_{\nearrow}(\alpha)$ and the down operator $P^{(k)}_{\searrow}(\alpha)$, such that
\[
\left(P^{(k)}_{\nearrow}(\alpha)\right)_{rs} =
\left\{
	\begin{array}{cl}
	1 & \mbox{ if } r=s \mbox{ and } \forall\, 1\leq i\leq \#\alpha \mbox{ it holds } r,s\notin\{j_{i}^{\alpha}, j_{i}^{\alpha}+1\},\\
	\frac{1}{2} & \mbox{ if } \exists\, 1\leq i\leq \#\alpha \mbox{ such that } r,s\in\{j_{i}^{\alpha}, j_{i}^{\alpha}+1\},\\
	0 & \mbox{ otherwise,}
	\end{array}
\right.
\]
\[
\left(P^{(k)}_{\searrow}(\alpha)\right)_{rs} =
\left\{
	\begin{array}{cl}
	\frac{1}{2} & \mbox{ if } r=s \mbox{ and } \exists\, 1\leq i\leq \#\alpha \mbox{ such that } r,s\in\{j_{i}^{\alpha}, j_{i}^{\alpha}+1\},\\
	-\frac{1}{2} & \mbox{ if } r\neq s \mbox{ and } \exists\, 1\leq i\leq \#\alpha \mbox{ such that } r,s\in\{j_{i}^{\alpha}, j_{i}^{\alpha}+1\},\\
	0 & \mbox{ otherwise.}
	\end{array}
\right.
\]
By calling $e_j$ the boolean row vector such that $(e_j)_i=1\Leftrightarrow i=j$, we define $A[a_1,\ldots,a_k]=\langle \Sigma\cup\{\#\}, \pi_0^{(k)}, \{O_\sigma^{(k)}\}_{\sigma\in\Sigma\cup\{\#\}}, F^{(k)} \rangle$ as the \mon\ where
\begin{itemize}
	\item $\pi_0^{(k)}=e_1\in\C^{1\times(k+1)}$,
	\item for $\alpha\in S$, the associated projectors of $O_\alpha^{(k)}$ are $P^{(k)}_{\nearrow}(\alpha)$ and $P^{(k)}_{\searrow}(\alpha)$,
	\item with each $O_\sigma^{(k)}$ such that $\sigma\in\Sigma\setminus S$, we associate the projector $I_{(k+1)\times(k+1)}$,
	\item the projector of the accepting result of $O_\#^{(k)}$ is $(e_{k+1})^Te_{k+1}$, i.e. the $(k+1)\times(k+1)$ boolean matrix having a 1 only in the bottom right entry.
\end{itemize}
A careful analysis of the behavior of $A[a_1, \ldots, a_k]$ leads to the following
\begin{theorem}\label{th:PTtoMON}
The automaton $A[a_1, \ldots, a_k]$ recognizes $L[a_1, \ldots,
a_k]$ with cutpoint $\lambda = \frac{1}{2^{2k+1}}$ isolated by 
$\delta = \frac{1}{2^{2(k+1)}}$.
\end{theorem}
Since the class \lipt\ is the boolean closure of languages of the form $L[a_1, \ldots, a_k]$, and $\lmo(\Sigma)$ is a boolean algebra, Theorem \ref{th:PTtoMON} implies that all literally idempotent piecewise testable languages can be recognized by \mons.
%
%
%
The observations made up to this point imply our main result:
\begin{theorem}\label{CharacterizationResult}
$
\lmo(\Sigma) = V_{\Sigma}(\overline{\mathbf{J}}) = \lipt(\Sigma).
$
\end{theorem}
%
%
%
Theorem \ref{CharacterizationResult} allows us to prove the existence of a polynomial time algorithm for deciding $\lmo(\Sigma)$ membership:
\begin{theorem}\label{th:LMOalgo}
Given a regular language $L\in\Sigma^*$, the problem of determining whether $L\in\lmo(\Sigma)$ is decidable in time $O((|Q|+|\Sigma|)^2)$, where $|Q|$ is the size of the minimal deterministic automaton for $L$.
\end{theorem}
This algorithm first constructs the minimal deterministic automaton $A_L$ for $L$ in time $O(|Q|\log(|Q|))$ as shown in \cite{H71}. Then, in time $O(|Q|)$, it checks whether $L$ is literally idempotent by visiting all the vertices in the graph of $A_L$. Finally, it verifies whether $L$ is piecewise testable in time $O((|Q|+|\Sigma|)^2)$ with the technique shown in \cite{T01}. The fact that $\lmo(\Sigma)=\lipt(\Sigma)$ completes the proof.

\subsubsection*{Acknowledgements:} The authors wish to thank Alberto Bertoni for the stimulating discussions which lead to the results of this paper.

  \newpage
\appendix
\section*{\mons\ recognizing literal idempotent piecewise testable languages}
  
We introduce two elementary operators of orthogonal projection as follows: 
\begin{equation*}
	\begin{split}
		P_{\nearrow} :=
		\left[ 
		\begin{array}{c c}
		+\frac{1}{2} & +\frac{1}{2} \\
		+\frac{1}{2} & +\frac{1}{2} \\
		\end{array} 
		\right],
	& \ \ \ \ \ \ \ \ \ \ \ \ \ 
		P_{\searrow} := 
		\left[ \begin{array}{c c}
		+\frac{1}{2} & -\frac{1}{2} \\
		-\frac{1}{2} & +\frac{1}{2} \\
		\end{array} \right].
	\end{split}
\end{equation*}
It is not hard to see that the up-operator $P^{(k)}_{\nearrow}(\alpha)$ is the diagonal block matrix having $P_{\nearrow}$ in correspondence of the indexes $j_{i}^{\alpha}, j_{i}^{\alpha}+1$, for all $1\leq i\leq \#\alpha$, and the identity matrix in the rest of the diagonal. More formally
\begin{equation*}
	\begin{split}
		P^{(k)}_{\nearrow}(\alpha) =\ 
	&
		\begin{array}{c | c | c | c | c | c | c | c | c | c  }
			& \mbox{\fontsize{6}{8}\selectfont $1,\ldots$} & \mbox{\fontsize{6}{8}\selectfont $j^{(\alpha)}_1,j^{(\alpha)}_1+1$} & \mbox{\fontsize{6}{8}\selectfont $\ldots$} & \mbox{\fontsize{6}{8}\selectfont $j^{(\alpha)}_2, j^{(\alpha)}_2+1$} &\mbox{\fontsize{6}{8}\selectfont $\ldots$} & \mbox{\fontsize{6}{8}\selectfont $\ \ldots\ $} &\mbox{\fontsize{6}{8}\selectfont $\ldots$} & \mbox{\fontsize{6}{8}\selectfont $j^{(\alpha)}_{\#\alpha}, j^{(\alpha)}_{\#\alpha}+1$} &\mbox{\fontsize{6}{8}\selectfont $\ldots,k+1$}\\ \hline
			\mbox{\fontsize{6}{8}\selectfont $1,\ldots$} & I & & & & & & & & \\ \hline
			\mbox{\fontsize{6}{8}\selectfont $\begin{array}{c}j^{(\alpha)}_1\\ \\j^{(\alpha)}_1+1\end{array}$} & & P_{\nearrow} & & & & & & & \\ \hline
			\mbox{\fontsize{6}{8}\selectfont $\ldots$} & & & I & & & & & & \\ \hline
			\mbox{\fontsize{6}{8}\selectfont $\begin{array}{c}j^{(\alpha)}_2\\ \\j^{(\alpha)}_2+1\end{array}$} & & & & P_{\nearrow} & & & & & \\ \hline
			\mbox{\fontsize{6}{8}\selectfont $\ldots$} & & & & & I & & & & \\ \hline
			\mbox{\fontsize{6}{8}\selectfont $\begin{array}{c} \\ \vdots \\ \ \end{array}$} & & & & & & \begin{array}{c} \\ \ddots \\ \ \end{array} & & & \\ \hline
			\mbox{\fontsize{6}{8}\selectfont $\ldots$} & & & & & & & I & & \\ \hline
			\mbox{\fontsize{6}{8}\selectfont $\begin{array}{c}j^{(\alpha)}_{\#\alpha}\\ \\j^{(\alpha)}_{\#\alpha}+1\end{array}$} & & & & & & & & P_{\nearrow} & \\ \hline
			\mbox{\fontsize{6}{8}\selectfont $\ldots,k+1$} & & & & & & & & & I 
		\end{array}
	\end{split}
\end{equation*}

On the other hand, the down-operator $P^{(k)}_{\searrow}(\alpha)$ is the diagonal block matrix having $P_{\searrow}$ in correspondence of the indexes $j_{i}^{\alpha}, j_{i}^{\alpha}+1$, for all $1\leq i\leq \#\alpha$, and zero anywhere else. More formally
\begin{equation*}
	\begin{split}
		P^{(k)}_{\searrow}(\alpha) =\ 
	&
		\begin{array}{c | c | c | c | c | c | c | c | c | c  }
			& \mbox{\fontsize{6}{8}\selectfont $1,\ldots$} & \mbox{\fontsize{6}{8}\selectfont $j^{(\alpha)}_1,j^{(\alpha)}_1+1$} & \mbox{\fontsize{6}{8}\selectfont $\ldots$} & \mbox{\fontsize{6}{8}\selectfont $j^{(\alpha)}_2, j^{(\alpha)}_2+1$} &\mbox{\fontsize{6}{8}\selectfont $\ldots$} & \mbox{\fontsize{6}{8}\selectfont $\ \ldots\ $} &\mbox{\fontsize{6}{8}\selectfont $\ldots$} & \mbox{\fontsize{6}{8}\selectfont $j^{(\alpha)}_{\#\alpha}, j^{(\alpha)}_{\#\alpha}+1$} &\mbox{\fontsize{6}{8}\selectfont $\ldots,k+1$}\\ \hline
			\mbox{\fontsize{6}{8}\selectfont $1,\ldots$} & \0 & & & & & & & & \\ \hline
			\mbox{\fontsize{6}{8}\selectfont $\begin{array}{c}j^{(\alpha)}_1\\ \\j^{(\alpha)}_1+1\end{array}$} & & P_{\searrow} & & & & & & & \\ \hline
			\mbox{\fontsize{6}{8}\selectfont $\ldots$} & & & \0 & & & & & & \\ \hline
			\mbox{\fontsize{6}{8}\selectfont $\begin{array}{c}j^{(\alpha)}_2\\ \\j^{(\alpha)}_2+1\end{array}$} & & & & P_{\searrow} & & & & & \\ \hline
			\mbox{\fontsize{6}{8}\selectfont $\ldots$} & & & & & \0 & & & & \\ \hline
			\mbox{\fontsize{6}{8}\selectfont $\begin{array}{c} \\ \vdots \\ \ \end{array}$} & & & & & & \begin{array}{c} \\ \ddots \\ \ \end{array} & & & \\ \hline
			\mbox{\fontsize{6}{8}\selectfont $\ldots$} & & & & & & & \0 & & \\ \hline
			\mbox{\fontsize{6}{8}\selectfont $\begin{array}{c}j^{(\alpha)}_{\#\alpha}\\ \\j^{(\alpha)}_{\#\alpha}+1\end{array}$} & & & & & & & & P_{\searrow} & \\ \hline
			\mbox{\fontsize{6}{8}\selectfont $\ldots,k+1$} & & & & & & & & & \0 
		\end{array}
	\end{split}
\end{equation*}

\end{document}